\documentclass[]{aiaa-tc}

\usepackage{bm}
\usepackage{stmaryrd}
\usepackage{graphicx}
\usepackage{wrapfig}
\usepackage[latin1]{inputenc}
\usepackage{times}
\usepackage{tikz}
\usepackage{amssymb,amsbsy,amsmath,amsfonts,amssymb,amscd}
\usepackage{verbatim}
\usepackage{algorithmic}
\usepackage{caption}
\usepackage{subcaption}

 \title{Implicit large-eddy simulation of compressible flows using the Interior Embedded Discontinuous Galerkin method}

 \author{
  P. Fernandez\thanks{Graduate Student, Department of Aeronautics and Astronautics, MIT, AIAA Student Member.} ,\ 
    N.C. Nguyen\thanks{Principal Research Scientist, Department of Aeronautics and Astronautics, MIT, AIAA Member.}\\
  {\normalsize\itshape
   Massachusetts Institute of Technology, Cambridge, MA 02139, United States}\\
  \and
  X. Roca%
   \thanks{Research Scientist, Barcelona Supercomputing Center.}\\
  {\normalsize\itshape
  Barcelona Supercomputing Center, Barcelona 08034, Spain}
  \and
  J. Peraire%
   \thanks{Professor and Department Head of Aeronautics and Astronautics, MIT, AIAA Fellow.}\\
  {\normalsize\itshape
  Massachusetts Institute of Technology, Cambridge, MA 02139, United States}
 }

 \AIAApapernumber{YEAR-NUMBER}
 \AIAAconference{Conference Name, Date, and Location}
 \AIAAcopyright{\AIAAcopyrightD{YEAR}}

\begin{document}

\maketitle

\begin{abstract}
We present a high-order implicit large-eddy simulation (ILES) approach for simulating transitional turbulent flows. The approach consists of an Interior Embedded Discontinuous Galerkin (IEDG) method for the discretization of the compressible Navier-Stokes equations and a parallel preconditioned Newton-GMRES solver for the resulting nonlinear system of equations. The IEDG method arises from the marriage of the Embedded Discontinuous Galerkin (EDG) method and the Hybridizable Discontinuous Galerkin (HDG) method. As such, the IEDG method inherits the advantages of both the EDG method and the HDG method to make itself well-suited for turbulence simulations. We propose a minimal residual Newton algorithm for solving the nonlinear system arising from the IEDG discretization of the Navier-Stokes equations. The preconditioned GMRES algorithm is based on a restricted additive Schwarz (RAS) preconditioner in conjunction with a block incomplete LU factorization at the subdomain level. The proposed approach is applied to the ILES of transitional turbulent flows over a NACA 65-(18)10 compressor cascade at Reynolds number 250,000 in both design and off-design conditions. The high-order ILES results show good agreement with a subgrid-scale LES model discretized with a second-order finite volume code while using significantly less degrees of freedom. This work shows that high-order accuracy is key for predicting transitional turbulent flows without a SGS model.

\end{abstract}



\section{Introduction}


Complete information about turbulent flows can be obtained by means of direct numerical simulation (DNS). Despite the availability of powerful supercomputers, DNS remains intractable for practical applications. Large-eddy simulation (LES) is a viable alternative to DNS. The central premise of LES is that large scales dominate the turbulent transport and energy budget, so that a numerical simulation will provide a realistic depiction of the flow if it captures those scales explicitly and somehow accounts for the small scales that cannot be resolved. Also, the small scales tend to be more homogeneous and isotropic and hence, are easier to model. Strategies for dealing with the small scales include explicit subgrid-scale (SGS) modeling and implicit numerical dissipation. In the classical (explicit) LES approach, the large-scale eddies of the flow field are resolved and the small scales are modeled using a SGS model. The development of SGS models has been a subject of intense interest for decades, in particular in the 1990s\cite{Lesieur:96, Meneveau:00}.




It turns out, however, that the leading-order truncation error term introduced by most numerical schemes is similar in form and magnitude to conventional SGS models\cite{Drikakis:07, Grinstein:07, Margolin:07, Fureby:99, Fureby:02, Rider:03}. A natural alternative to the classical LES approach is therefore to use the numerical dissipation of the discretization scheme to account for the dissipation that would take place in the unresolved scales, leading to the so-called implicit LES (ILES). Hence, the lack of a SGS model can be somehow compensated by either a finer mesh or a higher-order discretization, or both. Indeed, as pointed out by Boris\cite{Boris:90}, a factor of two increase in the spatial resolution will bring more improvement in the accuracy of the well-resolved scales than any SGS model would do. The ILES approach was first introduced in 1990 by Boris et al.\cite{Boris:90}, and has been successfully applied, among others, by Visbal et al.\cite{Galbraith:08, Visbal:09} using a compact difference method, and by Uranga et al.\cite{Uranga:11} using a Compact Discontinuous Galerkin (CDG) method\cite{Peraire:08}. ILES benefits from its easy implementation and computational robustness, and currently gains considerable attention from researchers in the computational fluid dynamics (CFD) community.

Second-order finite volume (FV) schemes have been widely used in academia and industry for the explicit LES of turbulent flows. However, they are not often used as an ILES solver because they tend to produce so much dissipation that even the large scales are not properly captured unless the mesh is extremely fine. Transitional turbulent flows share many features with other wave propagation phenomena for which high-order accuracy is known to be key. Let us consider natural transition to turbulence. The main difficulty to numerically capture transition is the very small magnitude of the perturbations than get exponentially amplified along the unstable portion of the laminar boundary layer. These small perturbations are ultimately responsible for the so-called non-linear breakdown and transition to turbulence. In particular, the amplitude of these unstable modes is several orders of magnitude below the freestream velocity at the location in which the boundary layer becomes unstable. As a result, very small amount of numerical dissipation and dispersion is needed to capture these instabilities and accurately predict the transition location. Over-dissipation may kill the small perturbations and lead to inaccurate transition prediction.

The  discontinuous Galerkin (DG)  method  has  many  of  the  advantages of  the continuous Galerkin (CG) method and the strengths of the FV method. In particular, it relies upon a strong mathematical  foundation  and  allows  for  high-order  implementations,  while  being  able  to handle  complex  geometries  and  mesh  adaptation.  In  addition, it  provides local conservation and stable discretization of the convective operator. On the other hand, the DG method requires significantly more operations per computational cell than the FV method. There is clearly a strong case to make DG methods more computationally efficient. In the spirit of making DG methods more competitive, researchers have developed more efficient DG schemes such as  the hybridizable DG (HDG) method\cite{Cockburn:093,Nguyen:091,Nguyen:092,Nguyen:11,Peraire:10} and the embedded DG (EDG) method\cite{Cockburn:093,Cockburn:092,Peraire:11}. Recently, a class of EDG methods\cite{Nguyen:15} were introduced as an effort to combine the best of the EDG and HDG schemes. One particular instance of this class of EDG methods is the interior EDG (IEDG) method\cite{Nguyen:15}. The IEDG method inherits many of the advantages of the EDG method and the HDG method, and is used in this paper for the ILES of turbulent flows. At present, turbulent flows simulated by using high-order DG methods are limited to Reynolds numbers of 100,000 or less. This paper is the first attempt at using a DG method for the ILES of aerodynamic turbulent flows with Reynolds number of 250,000.


This paper also focuses on the development of a parallel preconditioned Newton-GMRES solver for the nonlinear system of equations arising from the IEDG discretization of the Navier-Stokes equations. In particular, we introduce a minimal residual algorithm to compute a good initial guess for the Newton's method, and develop a restricted additive Schwarz (RAS) parallel preconditioner in conjunction with a block incomplete LU factorization at the subdomain level for the GMRES algorithm. The proposed approach is applied to the ILES of transitional turbulent flows over a NACA 65-(18)10 compressor cascade at Reynolds number 250,000 in both design and off-design conditions. The high-order ILES results show good agreement with the results of a subgrid-scale LES model discretized with a second-order FV code. Finally, we analyze the boundary layer structure and the transition mechanism in this type of flows. 

\section{Methodology}

\subsection{The Interior Embedded Discontinuous Galerkin method}

We consider the compressible Navier-Stokes equations written in dimensionless conservation form as
\begin{equation}
\label{NS}
\begin{array}{rcll}
\displaystyle \bm{q} - \nabla \bm{u}  & = & 0,  \quad \mbox{in } \Omega \times (0, T] , \\[1ex]
\displaystyle \frac{\partial  \bm{u}}{\partial t} +  \nabla  \cdot  \bm{F}(\bm{u},\bm{q}) - \bm s(\bm u, \bm q) & = & 0, \quad \mbox{in } \Omega \times (0, T] .
\end{array}
\end{equation}
Here, $\bm{u} = (\rho, \rho v_{j}, \rho E), \ j=1,...,d$ is the $m$-dimensional vector of dimensionless conserved quantities, $\bm{F}(\bm{u},\bm{q})$ are the Navier-Stokes fluxes of dimension $m \times d$
\begin{equation}
\bm{F}(\bm{u},\bm{q}) = \left( \begin{array}{c}
\rho v_j \\
\rho v_i v_j + \delta_{ij} p \\
 v_j (\rho E + p)
\end{array}
\right) - \left( \begin{array}{c}
0 \\
\tau_{ij}  \\
v_i \tau_{ij} + f_j
\end{array}
\right) ,
\end{equation}
and $\bm{s}(\bm{u},\bm{q})$ is a source term. For a calorically perfect gas in thermodynamic equilibrium, the viscous stress tensor, the heat flux, and the pressure are given by 
\begin{equation}
\tau_{ij} = \frac{1}{Re} \bigg[ \Big( \frac{\partial v_i}{\partial x_j}+\frac{\partial v_j}{\partial x_i} \Big) -\frac{2}{3}\frac{\partial v_k}{\partial x_k}\delta_{ij} \bigg] ,  \quad f_j = - \frac{\gamma}{Re \ Pr} \ \frac{\partial T}{\partial x_j} , \quad p = (\gamma - 1) \ \rho \ \Big( E-\frac{1}{2}v_k \ v_k \Big),
\end{equation}
where $Re$ is the Reynolds number, $Pr$ the Prandtl number, and $\gamma$ the specific heat ratio. Note that $\gamma = 1.4$ and $Pr = 0.72$ for air.  The freestream Mach number $M_\infty$ enters in the Navier-Stokes system through the dimensionless freestream pressure $p_{\infty} = 1 / (\gamma M^2_{\infty})$ since $\rho_\infty = U_\infty = 1$.  

The Interior Embedded Discontinuous Galerkin (IEDG) method is first introduced in the paper \cite{Nguyen:15}. The IEDG discretization\cite{Nguyen:15} of the compressible Navier-Stokes equations (\ref{NS}) reads as follows: Find $(\bm{q}_h,\bm{u}_h, \widehat{\bm{u}}_h) \in \bm{\mathcal{Q}}_h^k \times \bm{\mathcal{V}}_h^k \times \bm{\mathcal{M}}_h^k$ such that
\begin{subequations}
\label{IEDG}
\begin{alignat}{2}
\big( \bm{q}_h, \bm{r} \big) _{\mathcal{T}_h} + \big( \bm{u}_h, \nabla \cdot \bm{r} \big)  _{\mathcal{T}_h} -  \big< \widehat{\bm{u}}_h, \bm{r} \cdot \bm{n} \big> _{\partial \mathcal{T}_h}  & =  0, \\
\Big( \frac{\partial \bm{u}_h}{\partial t}, \bm{w} \Big)_{\mathcal{T}_h} - \Big( \bm{F}(\bm{u}_h,\bm q_h) , \nabla \bm{w} \Big) _{\mathcal{T}_h}  +  \left\langle \widehat{\bm{f}}_h(\widehat{\bm{u}}_h,\bm{u}_h,\bm{q}_h), \bm{w} \right\rangle_{\partial \mathcal{T}_h}  & =  \Big( \bm{s}(\bm{u}_h,\bm{q}_h), \bm{w} \Big)_{\mathcal{T}_h},  \\
\left\langle \widehat{\bm{f}}_h(\widehat{\bm{u}}_h,\bm{u}_h,\bm{q}_h), \bm{\mu} \right\rangle_{\partial \mathcal{T}_h \backslash \partial \Omega} + \left\langle \widehat{\bm{b}}_h(\widehat{\bm{u}}_h,\bm{u}_h,\bm{q}_h), \bm{\mu} \right\rangle_{\partial \Omega} =  0 , 
\end{alignat}
for all $(\bm{r},\bm{w}, {\bm{\mu}}) \in \bm{\mathcal{Q}}^k_h \times \bm{\mathcal{V}}^k_h \times \bm{\mathcal{M}}_{h}^k$. The numerical flux is defined as
\begin{equation}
\widehat{\bm{f}}_h = \bm{F}(\widehat{\bm{u}}_h,\bm{q}_h) \cdot \bm{n} + \bm{S} \cdot (\bm{u}_h - \widehat{\bm{u}}_h), \quad \mbox{on } \partial \mathcal{T}_h ,
\end{equation}
\end{subequations}
and  the boundary flux $\widehat{\bm{b}}_h$ depends on the boundary conditions. Here $\bm S$ denotes the stabilization tensor and is calculated using either the Lax-Friedrich method or the Roe's method \cite{Nguyen:093}.  We refer to \cite{Nguyen:093} for the definition of the inner products $(\cdot, \cdot)_{\mathcal{T}_h}$ and $\left\langle \cdot, \cdot \right\rangle_{\partial \mathcal{T}_h}$, as well as the approximation spaces $\bm{\mathcal{Q}}_h^k$, $\bm{\mathcal{V}}_h^k$ and $\bm{\mathcal{M}}_h^k$. We simply point out here that, in the IEDG method, $\bm{\mathcal{M}}_h^k$ consists of face-wise polynomials that are continuous for the interior faces and discontinuous for the boundary faces. Also, the superscript $k$ denotes the polynomial degree, while the subscript $h$ denotes the mesh size. 

At the inlet and outlet sections of the flowfield, the boundary flux $\widehat{\bm{b}}_h$ is defined as
\begin{equation}
\widehat{\bm{b}}_h = \frac{1}{2} (\bm{A}_n + |\bm{A}_n|) \cdot (\widehat{\bm{u}}_h - \bm{u}_h) + \frac{1}{2} (\bm{A}_n - |\bm{A}_n|) \cdot (\widehat{\bm{u}}_h - \bm{u}_{b}) ,
\end{equation} 
where $\bm{A}_n$ denotes the Jacobian of the inviscid flux normal to the boundary, and $\bm{u}_{b}$ is the given boundary state. Non-slip, adiabatic wall boundary condition is used on the airfoil surface, and periodicity is imposed on some portion of the boundary as necessary.

The semi-discrete system (\ref{IEDG}) is further discretized in time using diagonally implicit Runge-Kutta (DIRK) schemes \cite{Alexander77}. This results in a nonlinear system of equations whose solution will be described next.

\subsection{Solution of the numerical discretization}

\subsubsection{Minimal residual Newton algorithm}

Let us denote $\bm z_h = (\bm q_h, \bm u_h) \in \bm{\mathcal{Q}}_{h}^k \times \bm{\mathcal{V}}_{h}^k$. At any given time step $n$, the nonlinear system of equations that arises from the temporal discretization of the semi-discrete form (\ref{IEDG}) can be written as
\begin{subequations}
\label{NLS}
\begin{alignat}{2}
\bm R_{\rm NS}(\bm z_h^n,  \widehat{\bm u}_h^n) & = \bm{0}, \\[1ex]
\bm R_{\rm FL}(\bm z_h^n,  \widehat{\bm u}_h^n) & = \bm{0},
\end{alignat}
\end{subequations}
where $\bm R_{\rm NS}$ and $\bm R_{\rm FL}$ are the discrete non-linear residuals associated with  (\ref{IEDG}a)-(\ref{IEDG}b) and  (\ref{IEDG}c), respectively. The Newton-Raphson method is used to solve the nonlinear system (\ref{NLS}). The convergence of Newton's method depends on an initial guess $(\bm z_h^{n,0},  \widehat{\bm u}_h^{n,0})$.  We propose a minimal residual (MR) algorithm to compute the initial guess. In particular, we express the initial guess as a linear combination of the solutions at $s$ previous time steps as $(\bm z_h^{n,0},  \widehat{\bm u}_h^{n,0})  = \sum_{j = 1}^{s} \alpha_j  (\bm z_h^{n-j},  \widehat{\bm u}_h^{n-j})$, where the coefficients $\alpha_j$ are found as the minimizer of the following nonlinear least squares problem
\begin{equation}
\label{MinRes2}
(\alpha_1,\ldots,\alpha_s)  = \arg \min_{(\beta_1,\ldots,\beta_s)  \in \mathbb{R}^s}  \left\| \bm R_{\rm NS}\Big( \sum_{j = 1}^{s} \beta_j  (\bm z_h^{n-j},  \widehat{\bm u}_h^{n-j}) \Big) \right\|^2 + \left\| \bm R_{\rm FL}\Big( \sum_{j = 1}^{s} \beta_j  (\bm z_h^{n-j},  \widehat{\bm u}_h^{n-j}) \Big) \right\|^2 .
\end{equation}
We solve this optimization problem by using the Levenberg--Marquardt (LM) algorithm. The LM algorithm needs the gradient vectors $\partial \bm R / \partial \alpha_i$, which are computed approximately by finite difference. This in turns requires a small number of residual evaluations.



By applying Newton's method to (\ref{NLS}), we obtain a linear system of the form
\begin{equation}
\left[
\begin{array}{ccc}
\mathbf{A} &  \mathbf{B}  \\
\mathbf{C} & \mathbf{D} 
\end{array}
\right]
\left(
\begin{array}{c}
\delta \mathbf{Z} \\
\delta \widehat{\mathbf{U}}
\end{array}
\right)
=
\left(
\begin{array}{c}
\mathbf{F} \\
\mathbf{G}
\end{array}
\right) .
\end{equation}
Here, $\delta \mathbf{Z}$ is the vector of degrees of freedom for $\delta \bm z_h^n = (\delta \bm{q}_h^n, \delta \bm{u}_h^n) \in \bm{\mathcal{Q}}_{h}^k \times \bm{\mathcal{V}}_{h}^k$, and $\delta \widehat{\mathbf{U}}$ is the vector of degrees of freedom for $\delta \widehat{\bm{u}}_h^n  \in \bm{\mathcal{M}}_{h}^k$. We note that the matrix $\mathbf{A}$ has block-diagonal structure due to the discontinuous nature of the approximation spaces, and hence $\delta \mathbf{Z}$ can be readily eliminated to obtain a reduced system in terms of $\delta \widehat{\mathbf{U}}$
\begin{equation}
\label{LS}
\mathbf{K} \; \delta  \widehat{\mathbf{U}} = \mathbf{R} \ , 
\end{equation}
where $\mathbf{K}= \mathbf{D} - \mathbf{C} \ \mathbf{A}^{-1} \mathbf{B}$ and $\mathbf{R} = \mathbf{G} -  \mathbf{C} \  \mathbf{A}^{-1} \mathbf{F}$.

This is the global system to be solved at every Newton iteration. The advantage of the IEDG method lies in the fact that the size and number of non-zeros of $\mathbf{K}$ is much smaller than that of other discontinuous Galerkin methods, while retaining most of the robustness of the HDG scheme.

\subsubsection{Parallel preconditioned GMRES solver}

The linear system (\ref{LS}) is solved in parallel using the GMRES method\cite{sasc86} with iterative classical Gram-Schmidt (ICGS) orthogonalization. In order to accelerate GMRES convergence, a left preconditioner $\mathbf{M}^{-1}$ is used and the linear system (\ref{LS}) is replaced by
\begin{equation}
\label{LS1}
\mathbf{M}^{-1} \mathbf{K} \; \delta  \widehat{\mathbf{U}} = \mathbf{M}^{-1} \mathbf{R} \ , 
\end{equation}
Prior to describing our parallel preconditioner, the domain decomposition strategy is presented. While other finite volume and discontinuous Galerkin methods only require partitioning the elements $K$ in $\mathcal{T}_h$, the IEDG scheme also requires partitioning the high-order nodes on the faces, as they are the entities in the linear system (\ref{LS}). In particular, the METIS software \cite{Karypis:13} is used to minimize the edge-cut of the ``{\it high-order face node}"$-${\it to}$-$``{\it high order face node}" connectivity graph through a multiway $k$-partitioning algorithm\cite{Roca:13}. This yields a partition for both the elements and the high-order nodes on the faces. The overlapping subdomains are then constructed from the nonoverlapping subdomains, as described later. We note that the partition of the high-order nodes determines the amount of communication between processors and the amount of work that each processor performs. The element partition, however, plays a secondary role and is only relevant for I/O duties.

A restricted additive Schwarz (RAS)\cite{Cai:99} method with $\delta$-level overlap is used as parallel preconditioner; that is,
\begin{equation}
\mathbf{M}^{-1} = \mathbf{M}_{RAS_{\delta}}^{-1} := \sum_{i=1}^{N} \mathbf{R}_{i}^{0} \ \mathbf{K}_{i}^{-1} \ \mathbf{R}_{i}^{\delta} ,
\end{equation}
where $ \mathbf{K}_{i} = \mathbf{R}_{i}^{\delta} \ \mathbf{K} \ \mathbf{R}_{i}^{\delta}$ is the subdomain problem, $\mathbf{R}_{i}^{\beta}$ is the restriction operator onto the subspace associated to the nodes in the $\beta$-level overlap subdomain number $i$, and $N$ denotes the number of subdomains (i.e. processors). In the IEDG framework, $\delta$-level overlap corresponds to $\delta$-``high-order face node" overlap, as these are the entities that participate in the global problem. In other words, the $\delta$-level overlap subdomain includes all high-order face nodes in the nonoverlapping subdomain, as well as all the nodes in the $\delta$ neighboring layers of the ``{\it high-order face node}"$-${\it to}$-$``{\it high order face node}" connectivity graph. We note that this may result in different overlapping subdomains to those based on $\delta$-element overlapping. Also, the choice $\delta = 0$ leads to the so-called block Jacobi (BJ) preconditioner. From our experience, $\delta = 1$ provides the best balance between communication cost and number of GMRES iterations for the flow regimes considered in this paper, and is therefore employed. This result is consistent with the findings in \cite{Roca:13} for the HDG method.

In practice, the inverse of the subdomain problem is approximated by
\begin{equation}
\mathbf{K}_{i}^{-1} \approx \mathbf{\widetilde{U}}_{i}^{-1} \ \mathbf{\widetilde{L}}_{i}^{-1} ,
\end{equation}
where $\mathbf{\widetilde{L}}_{i}$ and $\mathbf{\widetilde{U}}_{i}$ denote the zero fill-in $(m \times m)-$block incomplete LU factors of $\mathbf{K}_{i}$. We shall refer to this as the BILU(0) factorization of $\mathbf{K}_{i}$. Also, a generalization for the IEDG method of the minimum discarded fill (MDF) reorder algorithm\cite{Persson:08} is used at the subdomain level.


Fast matrix-vector products and preconditioner solves are required for an efficient solution of the linear system (\ref{LS}). Here we take advantage of computer memory hierarchy and cache policies to that end. In particular, a block compressed row format is used to store $\mathbf{K}_{i}$ and its BILU(0) factors $\mathbf{\widetilde{L}}_{i}$ and $\mathbf{\widetilde{U}}_{i}$. The blocks are stored in memory in the same order as they will be accessed for the matrix-vector product and preconditioner solve, respectively. For the BILU(0) factors this is in turn determined by the MDF reorder. Also, the matrix-vector product is performed first for the interface nodes and then for the interior nodes in order to overlap communication and computation as much as possible.

\subsection{Boundary layer post-processing}

The ILES results will be post-processed to analyze the boundary layer (BL) structure. The treatment of the boundary layer is based on the pseudo-velocity, defined as
\begin{equation}
\bm{u}^*(\bm{s},n) := \int_0^n (\bm{\omega} \times \hat{\bm{n}}) \ dn \ , 
\end{equation}
instead of the actual velocity. Here, $\bm{\omega}$ denotes the vorticity and $(\bm{s},n)$ is the set of curvilinear coordinates associated to the airfoil surface. In particular, $\bm{s} = (s_1,s_2)$ and $n$ are the coordinates along the streamwise, cross-flow, and outward normal to the airfoil directions, respectively. Also, the unit vectors associated to these coordinates are denoted by $\hat{\bm{s}}_1$, $\hat{\bm{s}}_2$ and $\hat{\bm{n}}$, respectively. Unlike the actual velocity, the pseudo-velocity asymptotes to a constant outside the boundary layer, even with strong curvature, thus making the boundary layer edge a well-defined and systematically identifiable location. In particular, the edge of the BL, $n_{e}$, is computed as the first $n$ location simultaneously satisfying

\begin{equation}
\big| \bar{\bm{\omega}} \big| \ n < \epsilon_1 \big| \bar{\bm{u}}^* \big| , \quad \bigg| \frac{\partial{\bar{\bm{\omega}}}}{\partial{n}}\ \bigg| \ n^2 < \epsilon_2 \big| \bar{\bm{u}}^* \big| \ , 
\end{equation}
where $\epsilon_1 = 0.01$ and $\epsilon_2 = 0.1$ are some properly tuned constants for a systematic and robust detection of the BL edge. The local streamwise and cross-flow unit vectors are then defined as

\begin{equation}
\hat{\bm{s}}_1 := \bar{\bm{u}}_e / \bar{u}_e,  \quad \hat{\bm{s}}_2 := \hat{\bm{s}}_1 \times \hat{\bm{n}} \ ,
\end{equation}
where $\bm{u}_e = \bm{u}^*(n_e)$ denotes the velocity at the edge of the boundary layer, $u_e = ||\bm{u}_e||_2$ is the edge velocity magnitude, and the overbar denotes temporal and cross-flow averaging, that is,
\begin{equation}
\bar{\bm{u}}_e(s_1) := \frac{1}{T \cdot \Delta s_2} \int_{0}^{T} \int_{0}^{\Delta s_2} {\bm{u}}_e(\bm{s},t) \ ds_2 \ dt .
\end{equation}
We note that cross-flow averaging is {\it allowed}, and it corresponds to spanwise averaging, due to the 2D-like geometry and boundary conditions of the problem. Hence, it is used to accelerate the convergence of the statistics of the turbulent flow. The average streamwise and cross-flow velocity profiles are then given by

\begin{equation}
\bar{u}_1(s_1,n) = \bar{\bm{u}}^*(s_1,n) \cdot \hat{\bm{s}}_1(s_1,n) , \quad \bar{u}_2(s_1,n) = \bar{\bm{u}}^*(s_1,n) \cdot \hat{\bm{s}}_2(s_1,n) .
\end{equation}
The boundary layer streamwise displacement thickness, momentum thickness and shape parameter are defined as
\begin{equation}
\delta^{*}(s_1) := \int_0^{n_e} \Big( 1-\frac{\bar{u}_1}{\bar{u}_e} \Big) \ dn , \quad \theta(s_1) := \int_0^{n_e} \Big( 1-\frac{\bar{u}_1}{\bar{u}_e} \Big) \ \frac{\bar{u}_1}{\bar{u}_e} \ dn , \quad H(s_1) = \frac{\delta^*}{\theta} . 
\end{equation}
Also, the fluctuating streamwise velocity $u_1^{\prime}$ and the amplification factor of streamwise perturbations $N_1$ are defined as
\begin{equation}
u_1^{\prime} (\bm{s},t) := u_1(\bm{s},t) - \bar{u}_1(s_1,n), \qquad N_1(s_1) := \ln\Bigg({\frac{A_1(s_1)}{{A_1}_0}}\Bigg) \ , 
\end{equation}
where $A_1(s_1)$ denotes the amplitude of the disturbances at the boundary layer location $s_1$,
\begin{equation}
A_1(s_1) = \frac{1}{\bar{u}_e(s_1) \sqrt{n_e(s_1)}} \sqrt{\int_0^{n_e}  {\overline{{u_1^{\prime}}^2} \ dn}} \ , 
\end{equation}
and ${A_1}_0$ is some reference amplitude. We note that ${A_1}_0$ shifts $N_1(s_1)$ by a constant factor but does not affect its growth rate. Finally, the non-dimensional velocity and distance to the wall in the turbulent boundary layer are defined as
\begin{equation}
u^{+} = \frac{\bar{u}_1}{u_{\tau}} \ , \qquad  n^{+} = \frac{n}{l_{\tau}} \ 
\end{equation}
in the inner layer, and
\begin{equation}
\Delta u^{+} = \frac{\bar{u}_1 - \bar{u}_{e}}{u_{\tau}} \ , \qquad \eta = \frac{n}{\delta^{*}} \ 
\end{equation}
in the outer layer. Here, $u_{\tau} = \sqrt{\tau_{w} / \rho}$ denotes the shear velocity and $l_{\tau} = \nu / u_{\tau}$ is the wall unit length.

\section{Numerical results}

\subsection{Problem description}

The IEDG method is applied to the transitional turbulent flow over a NACA 65-(18)10 compressor cascade at Reynolds number $Re=250,000$ and Mach number $M=0.081$. The blade solidity is $\sigma = 1.0$ and the stagger angle $\xi = 28.3 \ \textnormal{deg.}$ We discretize the computational domain using curved hexahedral meshes. In particular, the 3D meshes are generated through extrusion of 2D quadrilateral meshes, and the polynomial degree $p$ of the parametric mapping from the reference element to the physical elements is set equal to the polynomial degree $k$ of the numerical approximation. The extrusion length is set to $0.1 \ c$, where $c$ denotes the airfoil chord. Non-slip, adiabatic wall boundary condition is used on the airfoil surface, while periodicity is imposed on the top/bottom surfaces and on the extrusion direction. The polynomial degree $k = 2$ is used in all the computations. A third-order, three-stage DIRK(3,3) method is used to integrate the nonlinear system (\ref{NLS}) in time with a non-dimensional time-step of $dt^*=dt \cdot U_{\infty}/c=0.005$. Hence, the resulting discretization is third-order accurate in space and time.

Both design and off-design conditions are considered. The relative angle between the freestream and the blade chord is set to $16.7 \ \textnormal{deg.}$ for the design condition and  $25.7 \ \textnormal{deg.}$ for the off-design condition. A computational mesh of 427,040 hehexahedra is used for the design condition, while a mesh of 909,360 hehexahedra is employed for the off-design condition. Our simulation results will be compared to experimental data\cite{Emery:58} and to a subgrid-scale LES model discretized using a second-order finite volume code and 31,000,000 elements\cite{Medic2015}.

\subsection{Design condition}


Fig. 1 (left) shows the time and spanwise average negative pressure coefficient obtained using the IEDG ILES approach and  the SGS-LES finite volume code\cite{Medic2015}, as well as the experimental data\cite{Emery:58}. The skin friction coefficient is shown on the right of Fig. 1. Since the separation and reattachment locations coincide on pressure and suction sides for this flow condition, they are indicated with the same vertical black line for both sides. The ILES and SGS-LES results match remarkably well, except on the transition location on the pressure side. This in turn induces a discrepancy in the reattachment location. We note that transition is very sensitive to a number of numerical artifacts, such as numerical dissipation and dispersion or inadequate treatment of inflow and outflow boundary conditions, and these are most likely responsible for the disagreement in the exact transition location. 

Particular emphasis is placed on analyzing the boundary layer structure. Since similar behavior is displayed the upper and lower sides, the discussion is limited here to the suction side boundary layer. The analysis is based on the high-order IEDG ILES results, as no SGS-LES finite volume results or experimental data are available for comparison. Figure 2 shows the streamwise displacement and momentum thicknesses (left), and the streamwise shape parameter (right) along the suction side. From these figures, the flow separates at $s_{1,sep} = 0.497$ due to the adverse pressure gradient and a laminar separation bubble (LSB) forms. As discussed later, the boundary layer becomes very unstable after separation and it eventually transitions to turbulence. Right after transition, the flow reattaches at $s_{1,reatt} = 0.714$ and remains attached all the way until the trailing edge. The laminar separation and turbulent reattachment on pressure and suction sides are also illustrated in Fig. 3 through the instantaneous and time-average velocity magnitude. 




Owing to the lack of bypass and forced transition mechanisms and the 2D-like nature of the flow, natural transition due to two-dimensional unstable modes is expected. This is numerically confirmed by the IEDG ILES results. The two-dimensional nature of transition is illustrated in Fig. 4 through the amplitude of the streamwise vs. cross-flow instabilities on the pressure (left) and suction (right) sides. Tollmien-Schlichting (TS) waves form before the boundary layer separates, and Kelvin-Helmholtz (KH) instabilities are ultimately responsible for transition after separation\footnote{The authors consider TS and KH waves to be different phenomena. In particular, we refer to the unstable modes of the Orr-Sommerfeld equation as TS modes if the boundary layer is attached, and as KH modes if the boundary layer is separated.}. The former are shown in Fig. 5 (left) for different boundary layer locations prior to separation. Indeed, the Gaussian-like shape corresponds to TS waves propagating and getting exponentially amplified along the boundary layer. In particular, Fig. 5 (left) shows the superposition of (1) TS waves and (2) pressure waves generated in the turbulent boundary layer of the blade at hand and the upper blade. The latter effect is responsible for the non-zero fluctuating velocity outside the boundary layer. The exponential growth rate of TS waves is shown through the streamwise amplification factor $N_{1}$ on the right of Fig. 5, and is consistent with linear stability theory. The box on the right figure indicates the region of the boundary layer in which the TS waves on the left figure are located. It is worth noting the small magnitude of the instabilities compared to the freestream velocity. This shows that very small amount of numerical dissipation is required for transition prediction. After separation, Tollmien-Schlichting waves lead into Kelvin-Helmholtz instabilities, as illustrated on the left of Fig. 6. KH instabilities produce very rapid vortex growth and are ultimately responsible for natural transition to turbulence in the separated shear layer.


The non-dimensional velocity profile at different locations along the turbulent portion of the boundary layer are displayed in Fig. 7 for the inner (left) and outer (right) layers. The high-order IEDG ILES results properly capture the viscous sublayer $u^{+} = n^{+}$; which extends from $n^{+} = 0$ to $n^{+} \approx 8$. A log-layer $u^{+} = (1 / \kappa) \ \textnormal{log} \ (n^{+}) + C^{+}$ is also observed from $n^{+} = 20$ to $n^{+} = 200$, where the exact extremes depend on the local Reynolds number and the pressure gradient. The numerical results fit well the experimentally measured value for the von Kármán constant $\kappa \approx 0.40$ (and $C^{+} = 5.5$) for moderate pressure gradients, while smaller values of $\kappa$ are observed for strong adverse pressure gradients. 




\subsection{Off-design condition}
The pressure coefficient computed with the IEDG ILES approach and the SGS-LES finite volume code, as well as the experimental data are shown in Fig. 8. The disagreement between the experimental data and the numerical results is thought to be due to the formation of a boundary layer on the spanwise walls during the wind tunnel experiments\cite{Emery:58}. This would reduce the effective cross sectional area, accelerate the flow and shift the pressure coefficient as in Fig. 8. Again, the IEDG ILES results and the FV SGS-LES results agree well except on the transition and reattachment locations.

As in design condition, the boundary layer undergoes laminar separation and turbulent reattachment on both suction and pressure sides. The separation and reattachment locations are  $x_{sep} = 0.199$ and $x_{reatt} = 0.384$ on the suction side, and $x_{sep} = 0.566$ and $x_{reatt} = 0.894$ on the pressure side. This is illustrated in Fig. 9 through the instantaneous (top) and time-average (bottom) velocity magnitude. It is worth noting the large size of the separation bubble on the pressure side. This is also clear from the streamwise displacement and momentum thicknesses (left) and the streamwise shape parameter (right) along the pressure side in Fig. 10. TS waves and KH instabilities are responsible for natural transition to turbulence in the off-design condition as well, as illustrated in Figures 6 (right), 11 and 12. Finally, Fig. 13 shows the non-dimensional velocity profile at different turbulent boundary layer locations for the inner (left) and outer (right) layers, and the Lagrangian coherent structures (LCSs) of the turbulent flow captured by the high-order implicit LES approach are shown in Fig. 14 through the Q-criterion isosurface colored by pressure.

\section{Conclusions}
We have presented a high-order Interior Embedded Discontinuous Galerkin (IEDG) method for the implicit large-eddy simulation of compressible flows, as well as a strategy for the parallel solution of the resulting discretization. In particular, we propose a Newton-Krylov method based on the GMRES iteration and a minimal residual algorithm for the initial guess of the nonlinear system. We described an overlapping domain decomposition for the IEDG method, and applied it to construct a restrictive additive Schwarz parallel preconditioner\cite{Cai:99}. Also, a block incomplete LU factorization with zero fill-in and a minimum discarded fill reorder algorithm\cite{Persson:08} are employed at the subdomain level.

The proposed methodology was applied to the implicit LES of a compressor cascade at Reynolds number 250,000 in design and off-design conditions. Despite using significantly less degrees of freedom, our high-order ILES results showed good agreement with a subgrid-scale LES model discretized using a second-order finite volume code. Emphasis was then placed on analyzing the boundary layer structure. Laminar separation and turbulent reattachment were observed on the suction and pressure sides for the two operating conditions considered. Tollmien-Schlichting (TS) and Kevin-Helmholtz (KH) instabilities were numerically detected and concluded to be responsible for natural transition to turbulence. Our work shows that high-order accuracy is key for predicting transitional turbulent flows without a SGS model.
 

\section*{Acknowledgments}

The first author would like to acknowledge ``la Caixa" Foundation for the Graduate Studies Fellowship that support his work. We also gratefully acknowledge Pratt \& Whitney and the Air Force Office of Scientific Research for partially supporting this effort. The work of the third author is supported by the European Comission through the Marie Sklodowska-Curie Actions (HiPerMeGaFlowS project). Finally, we would like to thank Prof. M. Drela, Dr. M. Sadeghi and H.-M. Shang for their useful comments and suggestions, and Dr. C. Hill for providing with the computing resources to perform some of the simulations presented in this paper.

\begin{figure}
  \label{fig:CpCf_design}
\centering
  \includegraphics[trim=70 55 70 10]{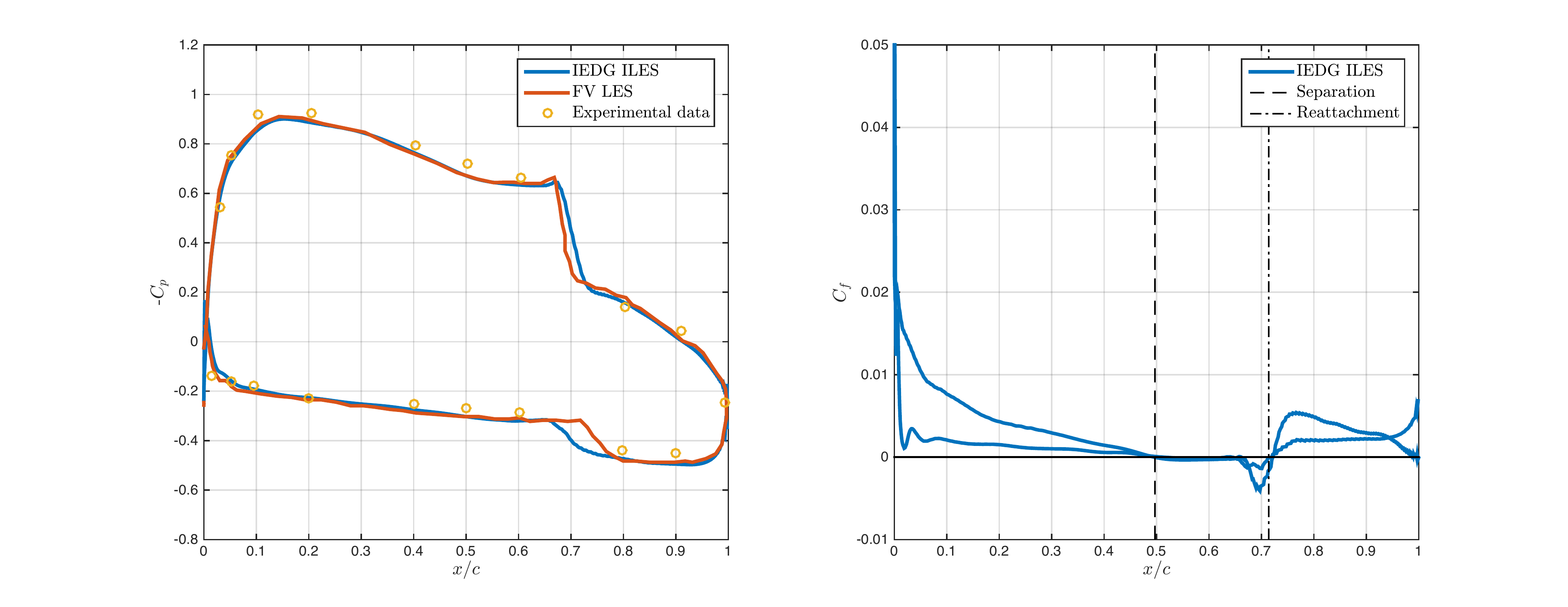}
 \captionsetup{justification=centering}
  \caption{Pressure coefficient (left) and skin friction coefficient (right) in design condition.}
\end{figure}

\begin{figure}
\centering
  \includegraphics[trim=70 55 70 10]{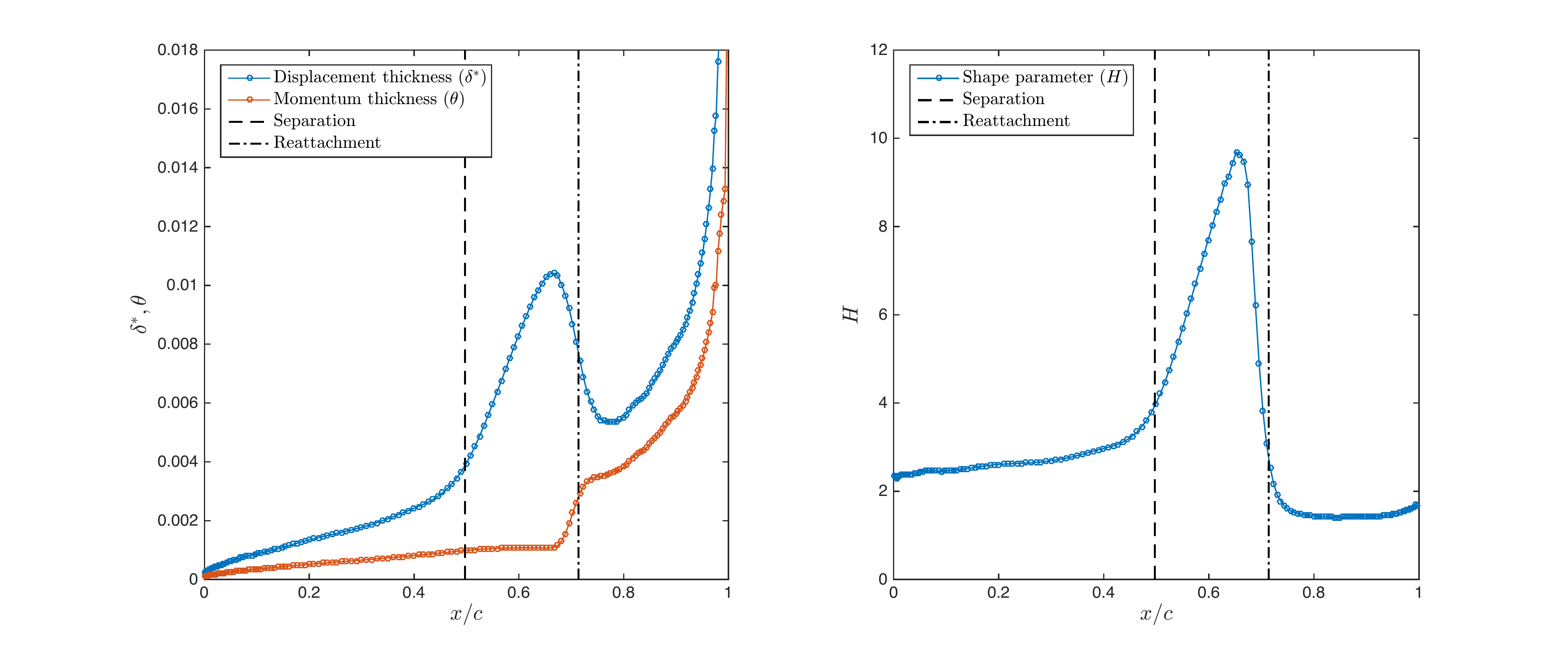}
  \label{fig:BLthicknessSuctionDesign}
 \captionsetup{justification=centering}
  \caption{Streamwise displacement thickness, momentum thickness and shape parameter on suction side and design condition.}
\end{figure}

\begin{figure}
\centering
  \includegraphics[trim= 0 60 0 0, width=105mm]{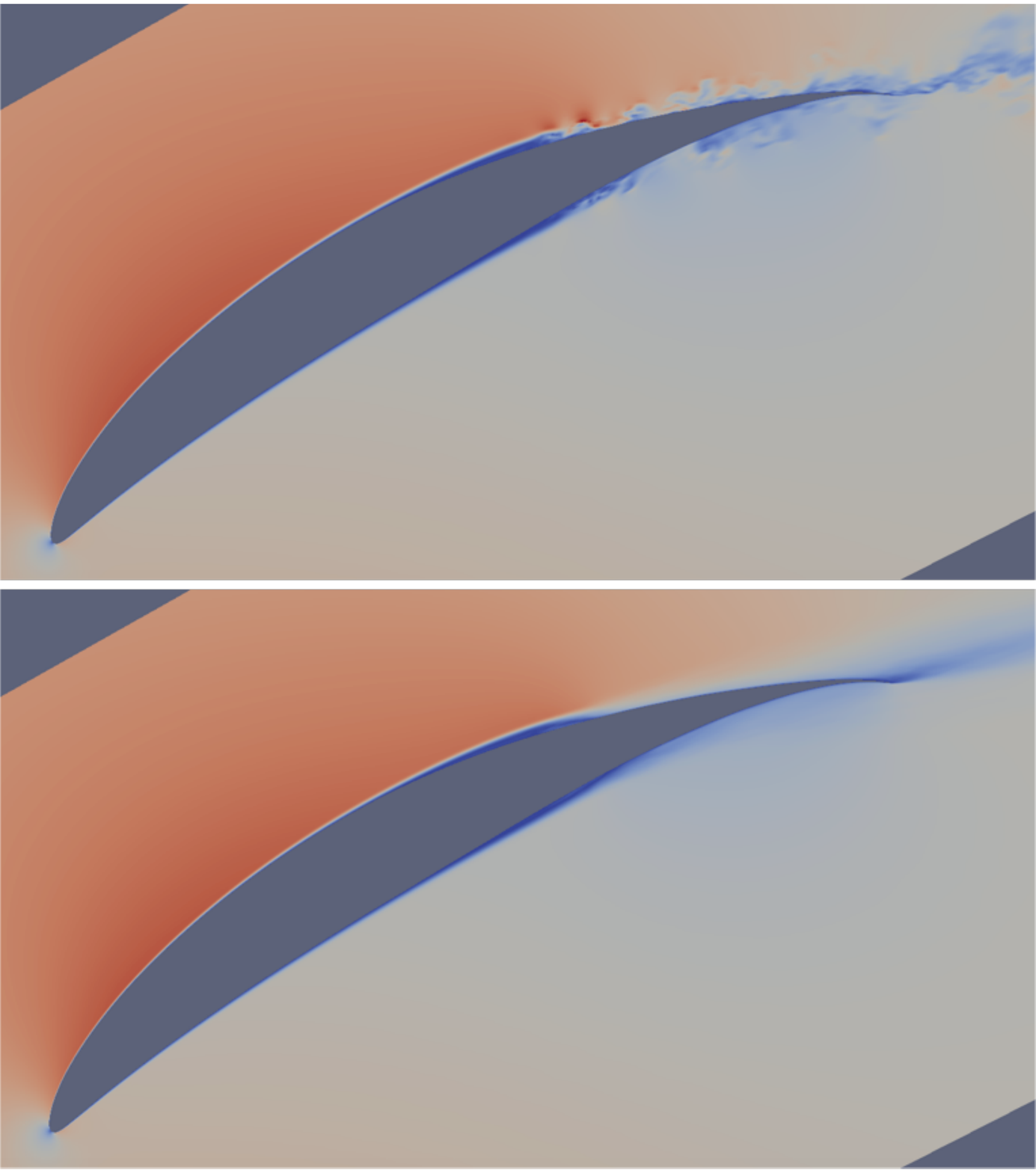}
  \label{fig:velMagDesign}
 \captionsetup{justification=centering}
  \caption{Instantaneous (top) and time-average (bottom) velocity magnitude in design condition.}
\end{figure}

\begin{figure}
\centering
  \includegraphics[trim=70 55 70 10]{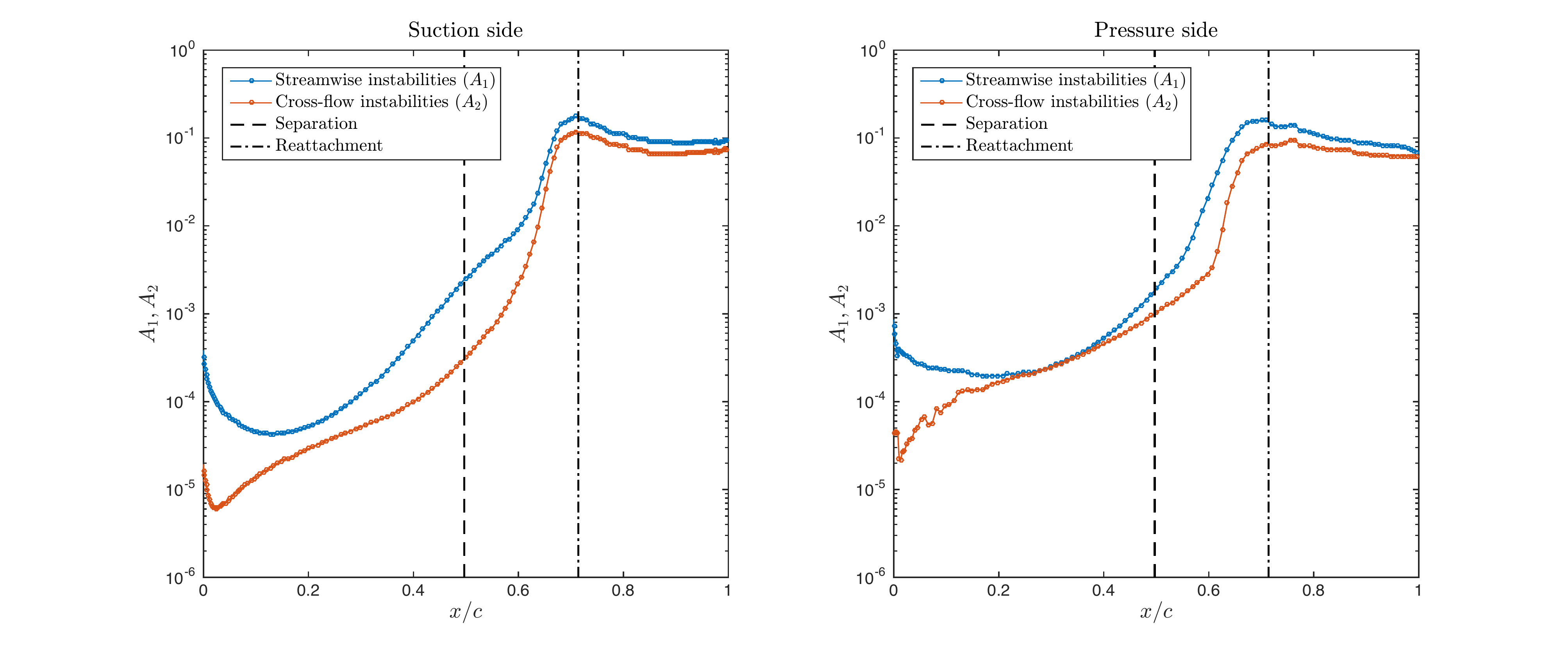}
  \label{fig:A1A2design}
 \captionsetup{justification=centering}
  \caption{Amplitude of streamwise and cross-flow instabilities on the pressure side (left) and the suction side (right) in design condition.}
\end{figure}

\begin{figure}
\centering
  \includegraphics[trim=70 55 70 10]{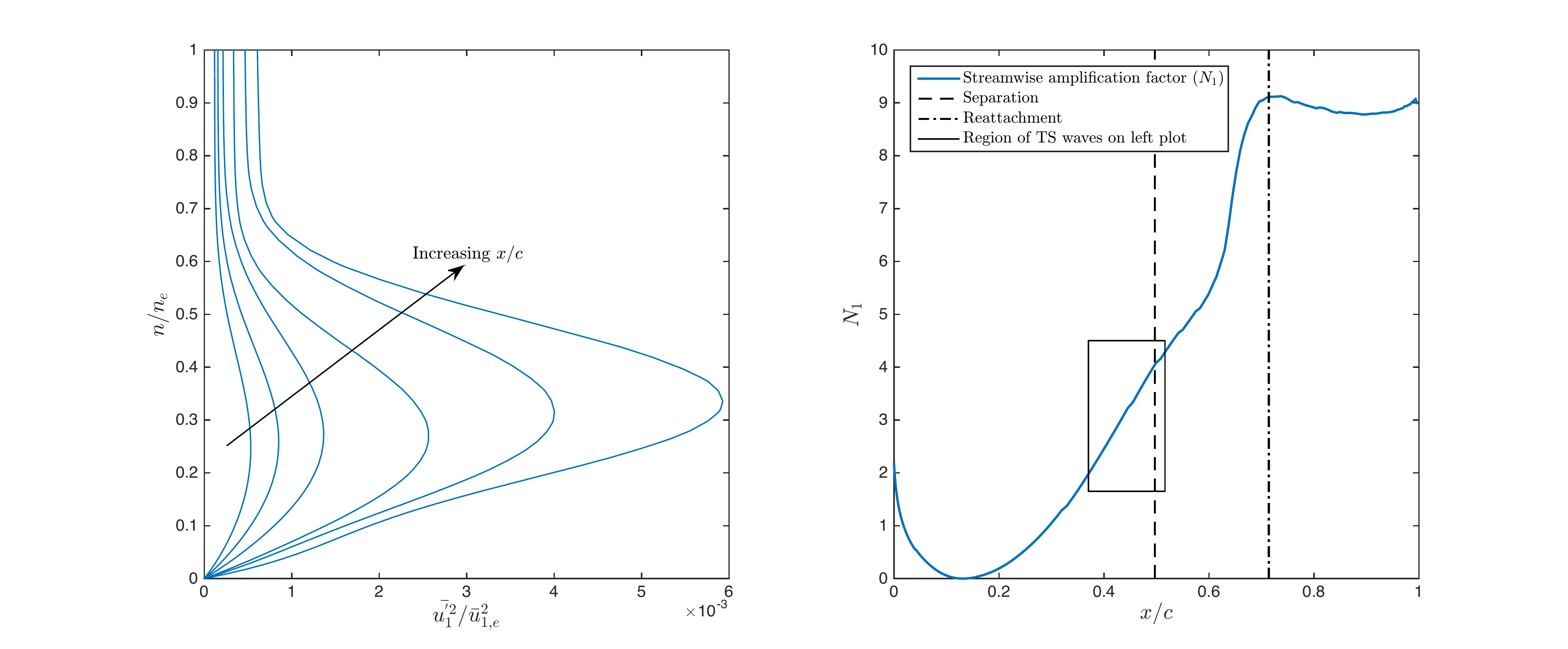}
  \label{fig:TSwavesSuctionDesign}
 \captionsetup{justification=centering}
  \caption{TS waves (left) and streamwise amplification factor (right) on suction side and design condition.}
\end{figure}

\begin{figure}
\centering
  \includegraphics[trim=70 55 70 10]{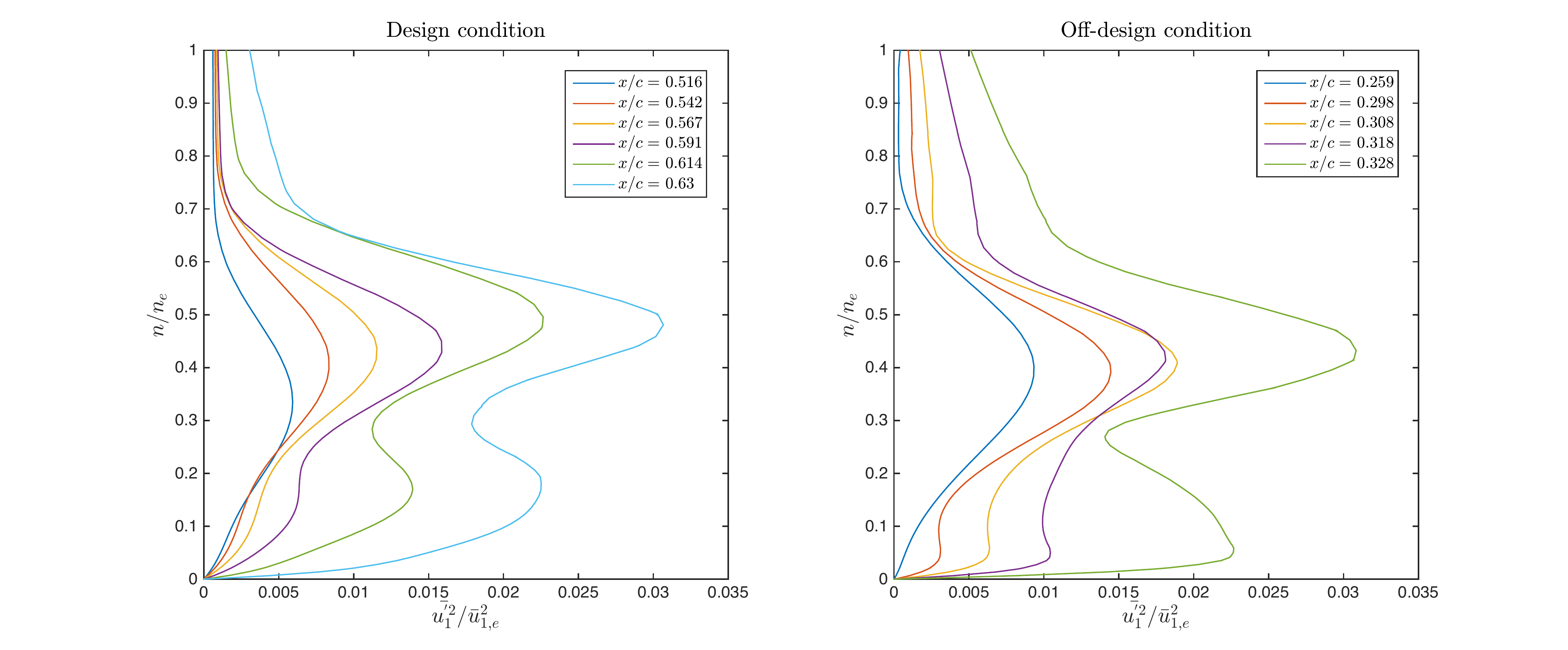}
  \label{fig:KHinstabilitiesSuction}
 \captionsetup{justification=centering}
  \caption{Transition from TS modes to KH modes along the separated, suction side boundary layer in design (left) and off-design (right) conditions.}
\end{figure}

\begin{figure}
\centering
  \includegraphics[trim=70 55 70 10]{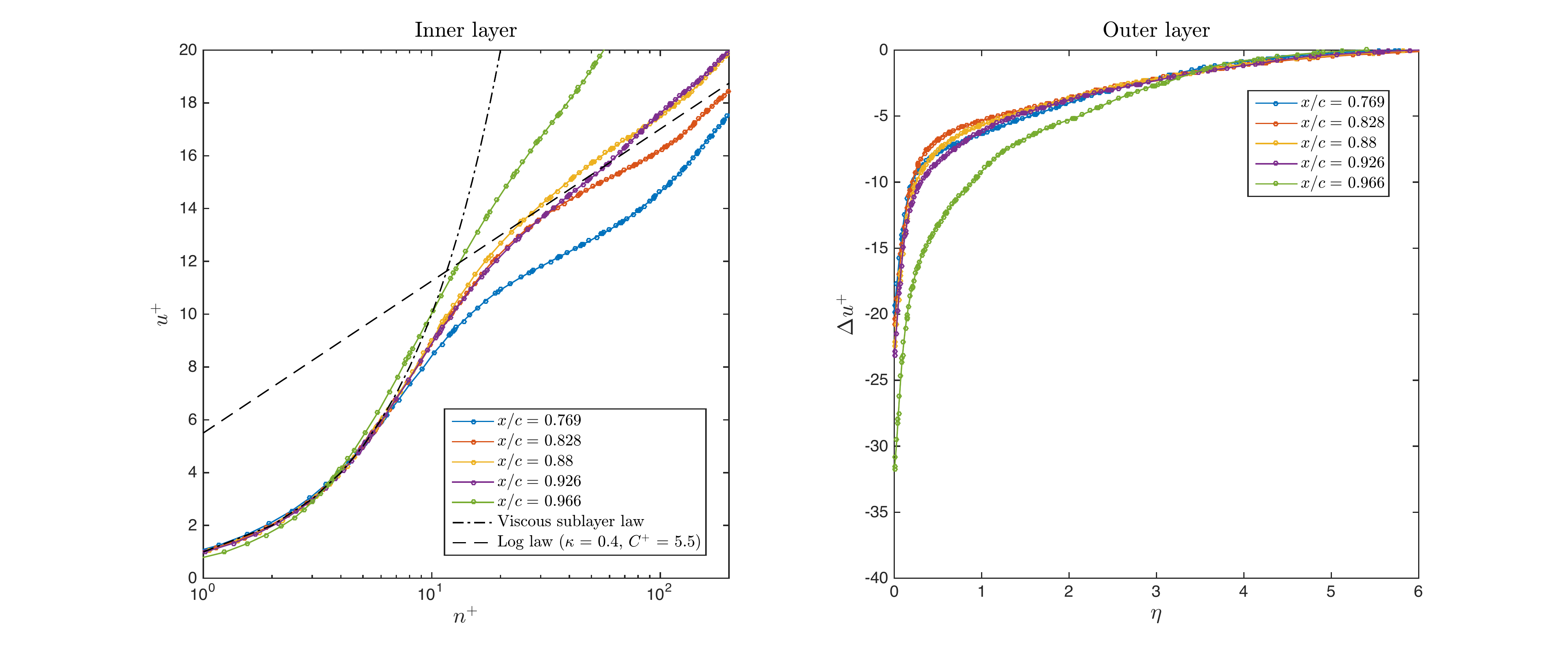}
  \label{fig:turbProfileSuctionDesign_v2}
 \captionsetup{justification=centering}
  \caption{Inner layer (left) and outer layer (right) non-dimensional velocity profiles at different locations of the suction side turbulent boundary layer in design condition.}
\end{figure}





\begin{figure}
\centering
  \includegraphics[trim=0 25 0 0, width=110mm]{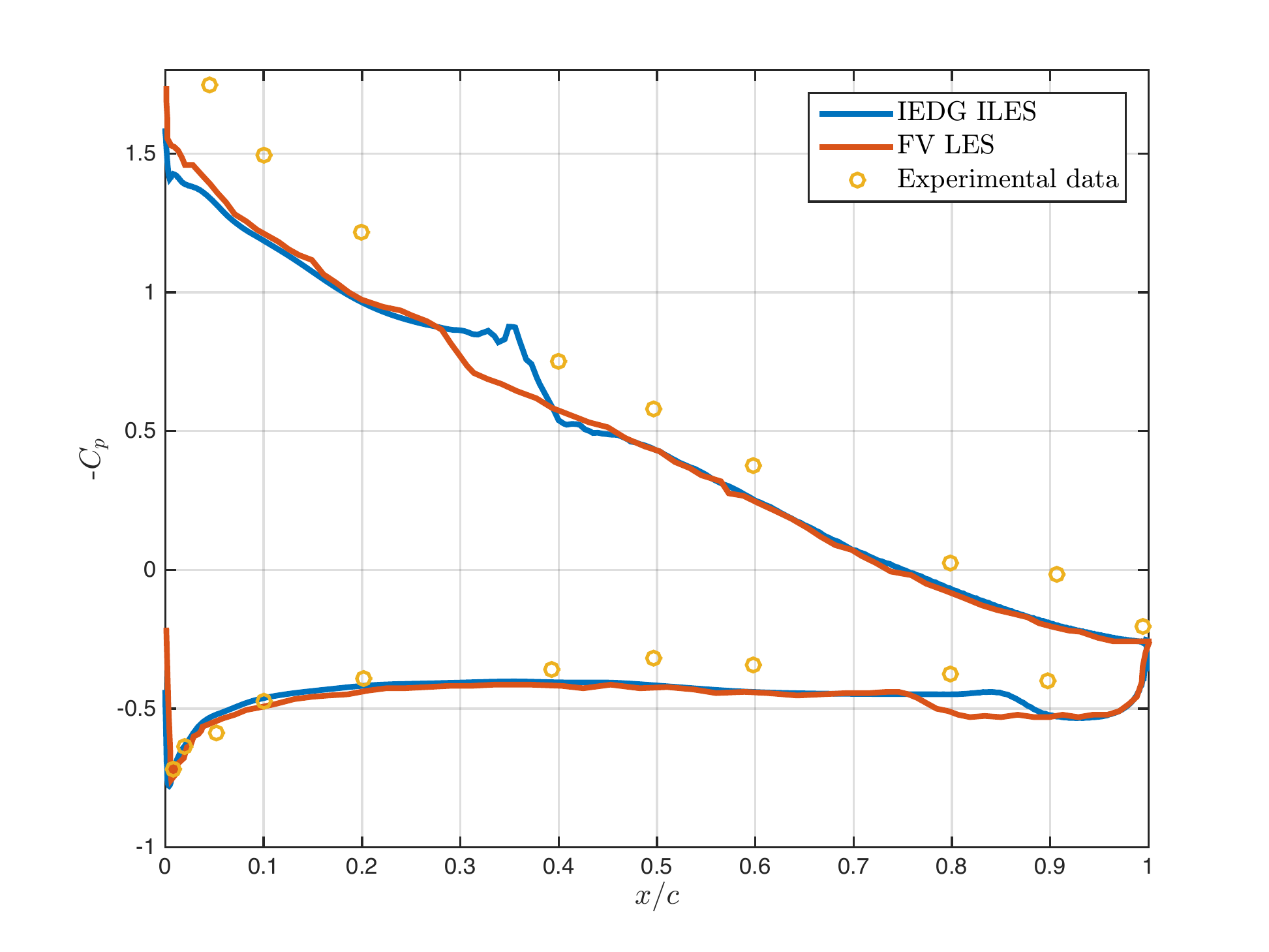}
  \label{fig:Cp_offDesign}
 \captionsetup{justification=centering}
  \caption{Pressure coefficient in off-design condition.}
\end{figure}

\begin{figure}
\centering
  \includegraphics[trim= 0 10 0 0, width=110mm]{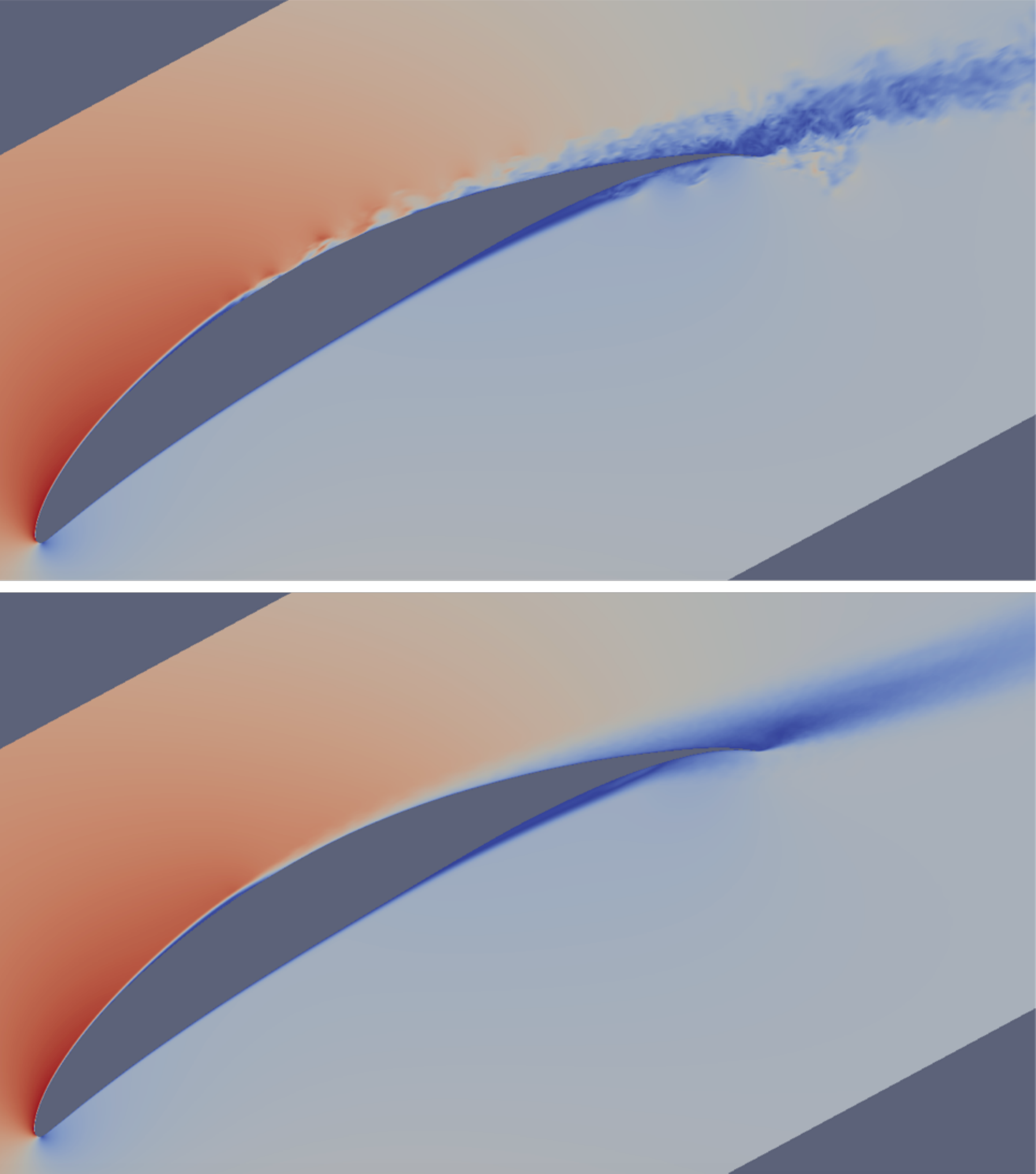}
  \label{fig:velMagOffDesign}
 \captionsetup{justification=centering}
  \caption{Instantaneous (top) and time-average (bottom) velocity magnitude in off-design condition.}
\end{figure}

\begin{figure}
\centering
  \includegraphics[trim=70 55 70 10]{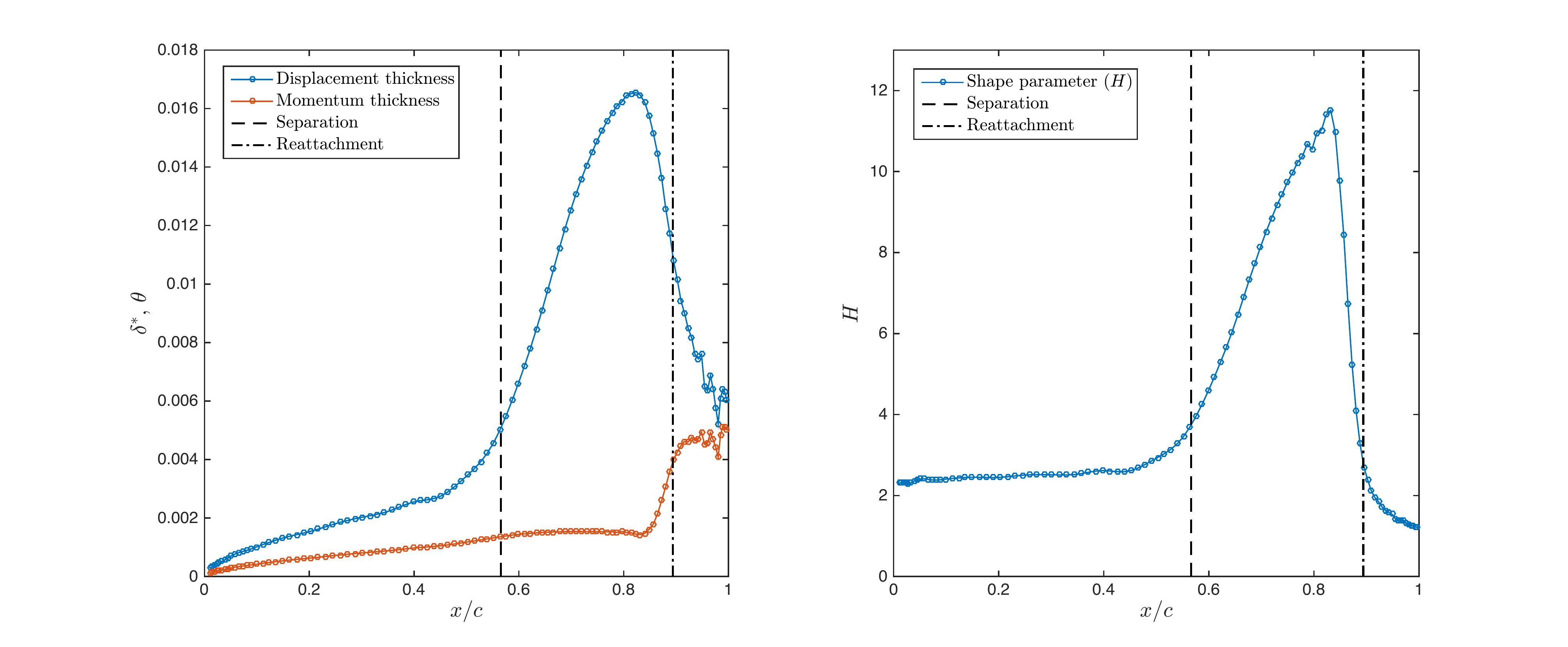}
  \label{fig:BLthicknessPressureOffDesign}
 \captionsetup{justification=centering}
  \caption{Streamwise displacement thickness, momentum thickness and shape parameter on pressure side and off-design condition.}
\end{figure}


\begin{figure}
\centering
  \includegraphics[trim=70 55 70 10]{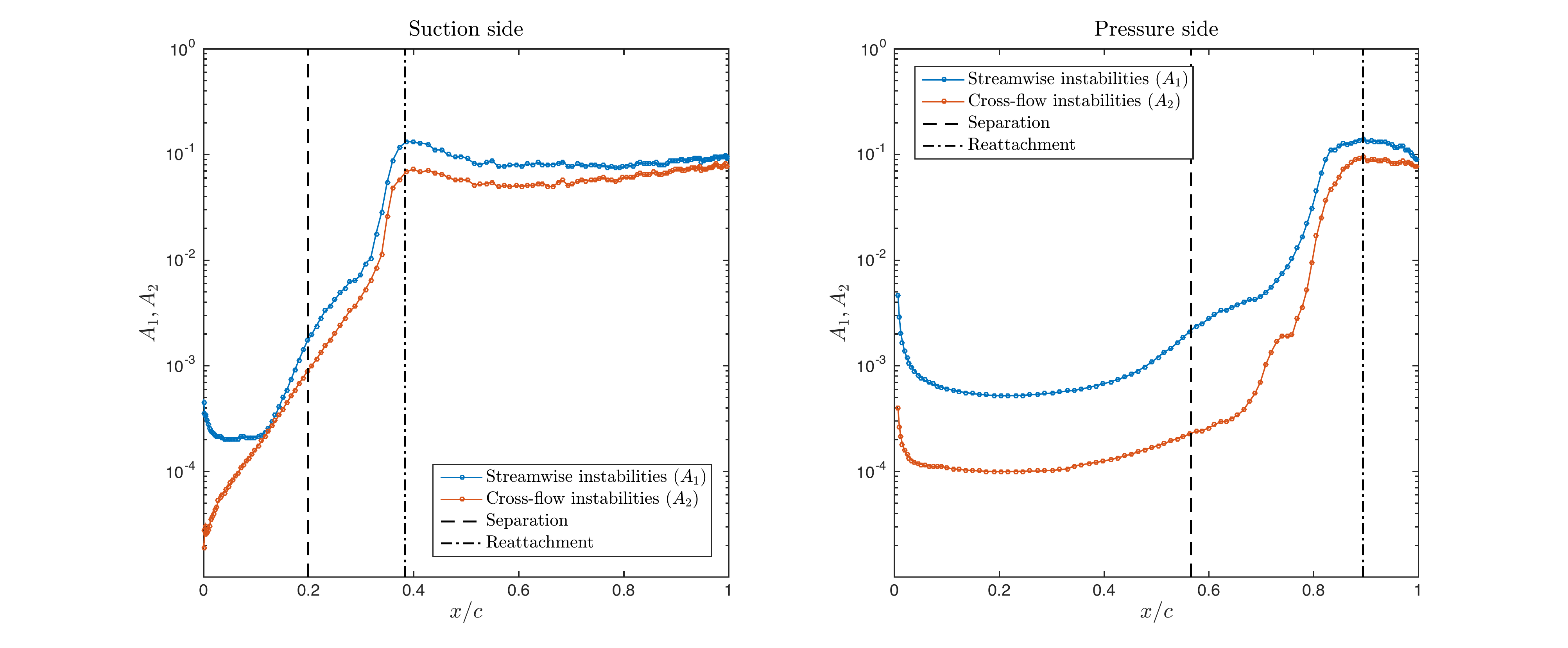}
  \label{fig:A1A2offDesign}
 \captionsetup{justification=centering}
  \caption{Amplification factor of streamwise and cross-flow instabilities on the pressure side (left) and the suction side (right) in off-design condition.}
\end{figure}

\begin{figure}
\centering
  \includegraphics[trim=70 55 70 10]{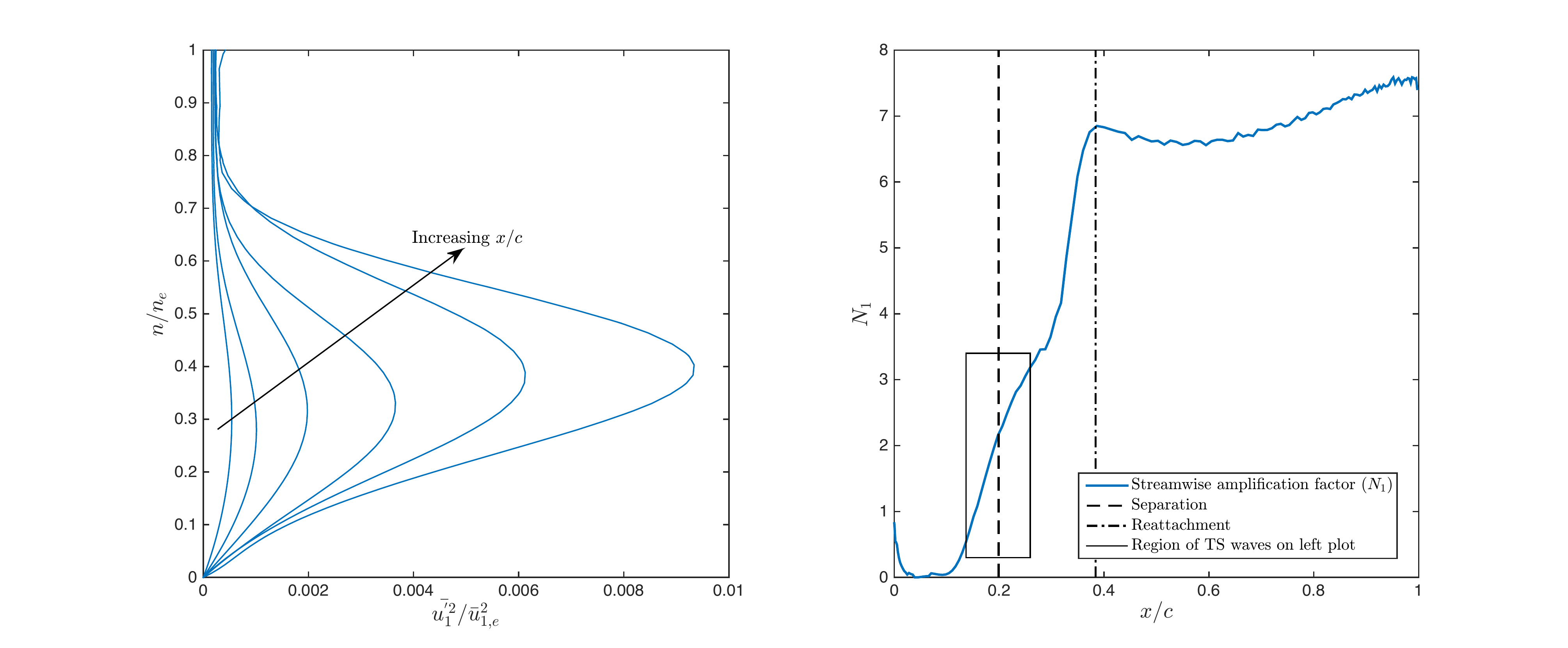}
  \label{fig:TSwavesSuctionOffDesign}
 \captionsetup{justification=centering}
  \caption{TS waves (left) and streamwise amplification factor (right) on the suction side boundary layer in off-design condition.}
\end{figure}

\begin{figure}
\centering
  \includegraphics[trim=70 55 70 10]{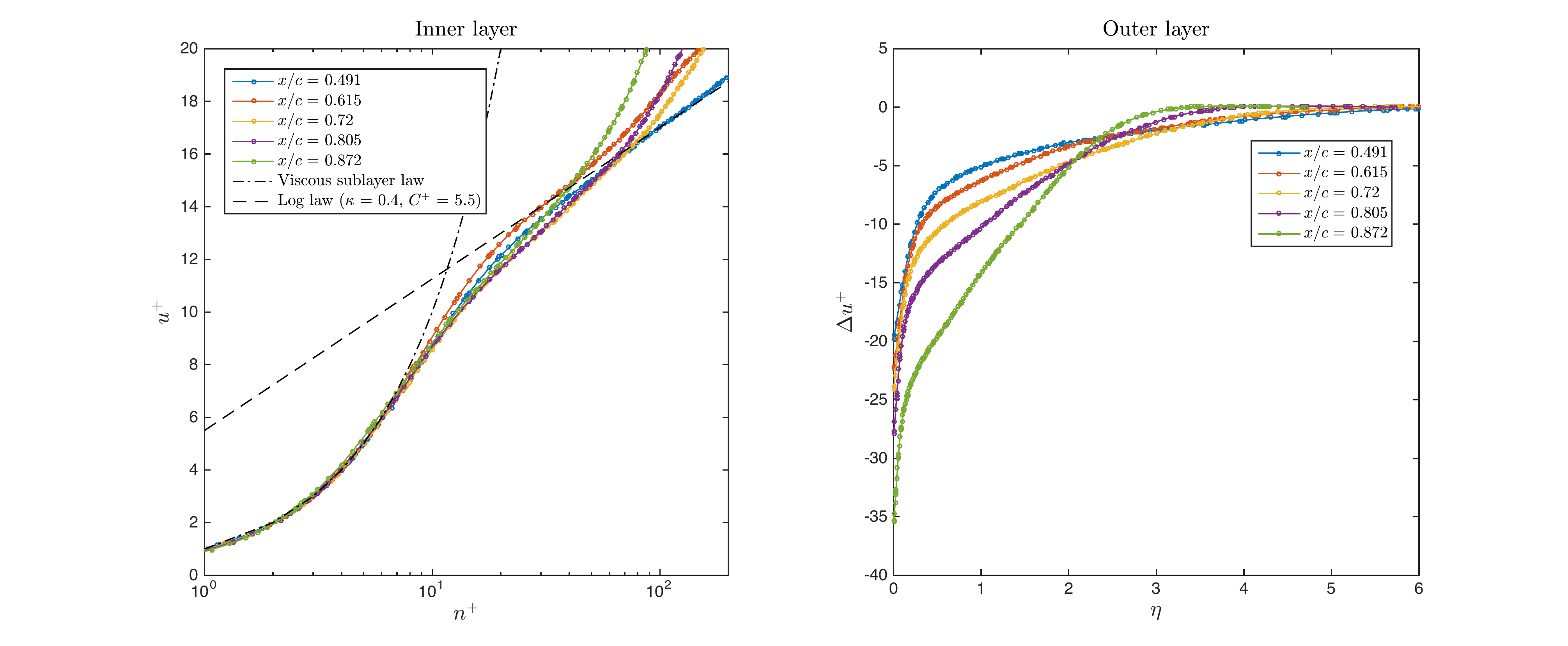}
  \label{fig:turbProfileSuctionOffDesign}
 \captionsetup{justification=centering}
  \caption{Inner layer (left) and outer layer (right) non-dimensional velocity profiles at different locations of the suction side turbulent boundary layer in off-design condition.}
\end{figure}

\begin{figure}
\centering
  \includegraphics[trim=0 65 0 0, width=105mm]{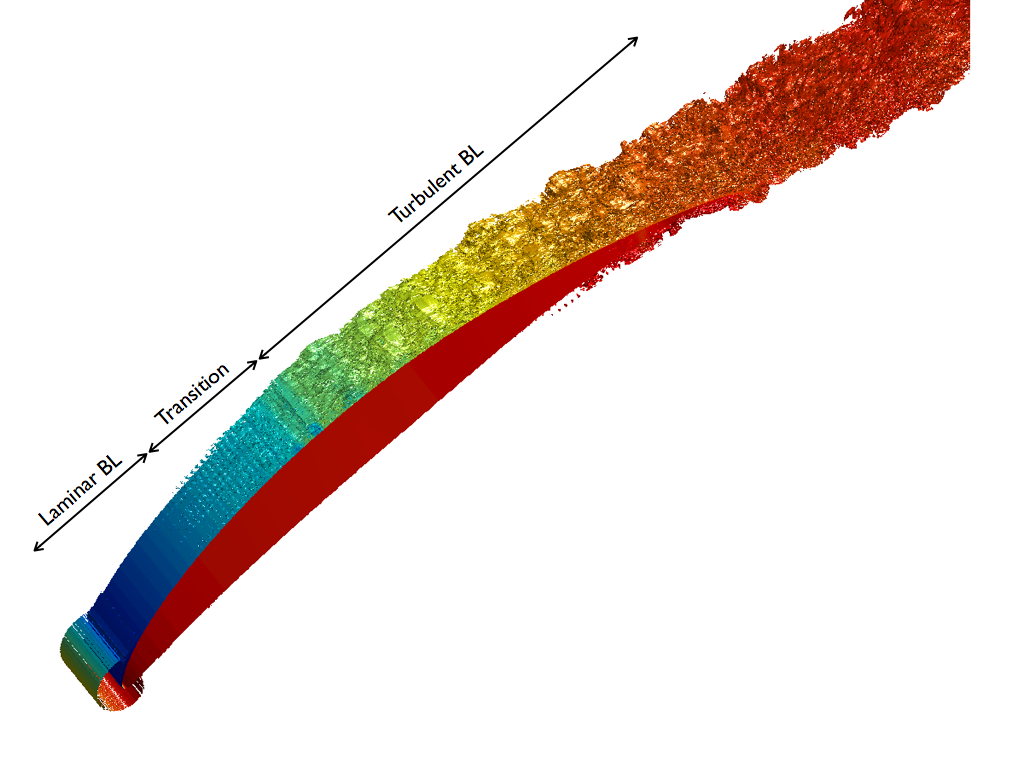}
  \label{fig:QcritOffDesign}
 \captionsetup{justification=centering}
  \caption{Instantaneous Q-criterion isosurface colored by pressure in off-design condition.}
\end{figure}


\end{document}